# Background Independent Quantum Mechanics, Classical Geometric Forms and Geometric Quantum Mechanics-II

Aalok*

Department of Physics, University of Rajasthan, Jaipur 302004, India;
and Jaipur Engineering College and Research Centre (JECRC), Jaipur 303905, India.

**Abstract**: The geometry of Quantum Mechanics in the context of uncertainty and complementarity, and probability is explored. We extend the discussion of geometry of uncertainty relations in wider perspective. Also, we discuss the geometry of probability in Quantum Mechanics and its interpretations. We give yet another interpretation to the notion of Faraday lines and loops as the *locus of probability flow*. Also, the possibilities of visualization of spectra of area operators by means of classical geometric forms and conventional Quantum Mechanics are explored.

___________________________________

*\*E-mail*: aalok@uniraj.ernet.in

# 1. Introduction

The discussion in this paper investigates the possible geometric consequences of the Background Independent Quantum Mechanics (BIQM) and an extended framework of Quantum Mechanics in comprehensive perspective as emphasized in recent papers [1, 2]. The present discussion echoes what Riemann suggested in his celebrated address [23] that the geometry of space may be more than just a fiducial mathematical entity serving as a passive stage for physical phenomena, and may in fact have a direct physical meaning in its own right.

The present paper addresses various aspects of geometry of Quantum Mechanics. Part of the underlying approach is based on the fact that uncertainty relation is at the centre-stage of the Background Independent formalism. Features of Quantum Mechanics such as uncertainties and state vector reductions can be reformulated geometrically [3]. The geometrical reformulation provides a unified framework to discuss number of issues including second quantization procedure, and the role of coherent states in semi-classical considerations, and to correct many a misconceptions [3].

Researchers studying gravity have also shown considerable interest in the geometric structures in quantum mechanics in general and projective Hilbert space in specific [3, 15-21]. In the light of recent studies [1, 2, 15, 16, 18-21] of geometry of the quantum state space, the need and call for further extension of standard geometric quantum mechanics is irresistible. Classical mechanics has deep roots in (symplectic) geometry while quantum mechanics is essentially algebraic. However, one can recast quantum mechanics in a geometric language, which brings out the similarities and differences between two theories [3]. The idea is to pass from the Hilbert space to the space of rays, which is the "true" space of states of quantum mechanics. As a quantum system evolves in time the state vector changes and it traces out a curve in the Hilbert space $\mathcal{H}$. Geometrically, the evolution is represented as a closed curve in the projective Hilbert space $\mathcal{P}$ [1, 2, 5-9, 12]. The space of rays- or the projective Hilbert space is in particular, a symplectic manifold which is equipped with a *Kähler* structure. Regarding it as a symplectic manifold, one can recast the familiar constructions of classical mechanics.

Deeper reflections show that Quantum Mechanics is in fact not as linear as it is advertised to be. The space of physical states is the space of rays in the $\mathcal{H}$, that is projective Hilbert space $\mathcal{P}$; and $\mathcal{P}$ is a genuine, non-linear manifold [3].

The organization of this paper is as follows: In the section 2 we continue the discussion of geometry of uncertainty relations in statistical and even wider perspective. In section 3, we discuss the geometry of probability and its interpretations. In the section 4, we explore the possibilities of visualization of spectra of area operators by means of classical geometric forms and using conventional Quantum Mechanics.

**2. Uncertainty and Complementarity**

We follow the consequences of the geometry of the uncertainty relation described in [1], wherein the geometry of the uncertainty relation $(\Delta p)(\Delta x) \approx \hbar$ is shown to be rectangular hyperbola. Which, in a way signifies the complementarity in Quantum Mechanics. Needless to say that similar geometrical attributes could be discussed in the context of uncertainty relation

$$(\Delta E)(\Delta t) \approx \hbar . \tag{1}$$

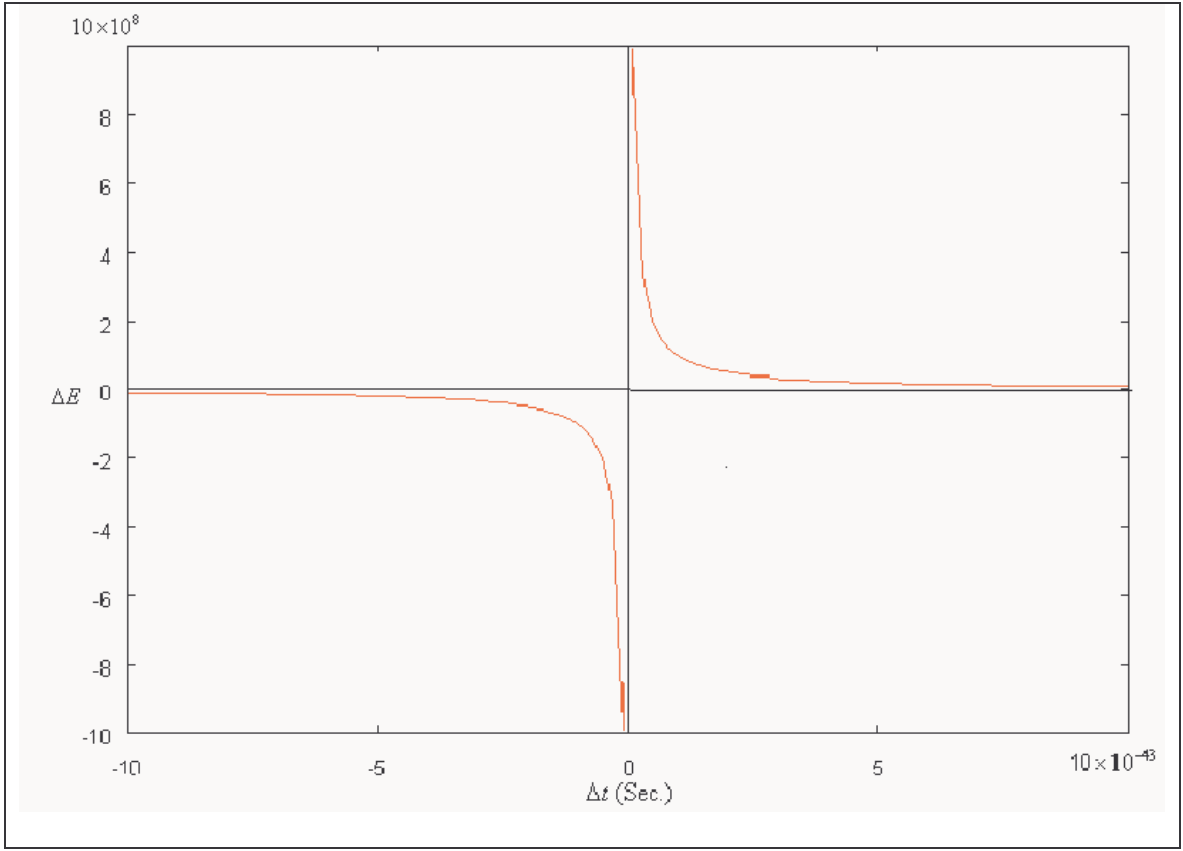

**Fig. 1:** The plot of the uncertainty relation $(\Delta E)(\Delta t) \approx Const$ (on arbitrary scale and axes could be interchanged), where $\Delta t$ can assume from negative finite to positive finite values.

But, we have a different point to be discussed in this regard. The variance $\Delta E$ could assume negative values as well. As the subject evolved with the statistical inputs, the meaning of the variance has broadened. This can be easily perceived. It is true that $(\Delta E)^2$ can not be negative, but, $(\Delta E)$ can certainly be positive as well as negative. We can discuss it as follow:

The variance $(\Delta E)^2$ is given as:

$$(\Delta E)^2 = <E^2> - <E>^2, \qquad (2)$$

and thus $(\Delta E) = \pm\sqrt{<E^2> - <E>^2}$. $\qquad (3)$

Which allows positive as well as negative values of $\Delta E$. The negative values of $\Delta E$ neither stand for negative energy nor do they imply that

$<E>^2 \geq <E^2>$.

The negative values of $\Delta E$ could be interpreted as variance about the mean or average energy. More importantly, the two curves in different quarters are mutually exclusive. And thus no trivial correlation could be established. Had it been in two adjacent quarters, then one could think of forging a correlation between two characteristic curves.

### 3. Probability and Geometry

It is worth mentioning here the importance of the geometrization of Quantum Mechanics discussed by Kibble [17], which pointed out that the Schrödinger evolution can be regarded as Hamiltonian flow on $\mathcal{H}$. We wish to recast the same spirit in terms of probability flow in $\mathcal{P}$ and in Probability space.

We examine the equation of continuity in the context of Quantum Mechanics as:

$$\frac{\partial P(x,t)}{\partial t} + (\vec{\nabla}.\vec{S}) = 0. \tag{4}$$

Where, $P = \Psi^*\Psi$, (5)

is the *probability density* and

$$\vec{S}(x,t) = -\frac{i\hbar}{2m}\left[\Psi^*\nabla\Psi - (\nabla\Psi^*)\Psi\right], \tag{6}$$

is *probability current density*.

Just as the equation of continuity says that no sources or sink of matter are present, this equation of continuity in Quantum Mechanics asserts that creation or destruction of probability that is any increase or decrease $(\partial P/\partial t)d\tau dt$ in the probability for finding the particle in a given volume element $d\tau \equiv d^3x$ is compensated by a corresponding decrease or increase elsewhere through an inflow or outflow of probability $(\text{Div } \vec{S})d\tau dt$ across the boundaries of $d\tau$.

We make the following observations and remarks in the context of equation of continuity (equation (4)). The term $\partial P/\partial t$ denotes rise and fall of probability with time, and thus depending on $(\partial P/\partial t)$, the divergence of probability current $\vec{S}$ could be interpreted differently as in the following discussion. The figure (2) shows a large *positive* divergence as the arrows are pointing in. The figure (3) shows a large *negative* divergence as the arrows are pointing out. The equation

$$(\text{Div } \vec{S}) = 0,$$

exhibits solenoidal nature of *probability flow* (in analogy with fluid dynamics) or *flow of the probability current*. This is true in case of magnetic field too, where, $\vec{\nabla}.\vec{B} = 0$ implies the Faraday lines form loops as shown in the figure (4).

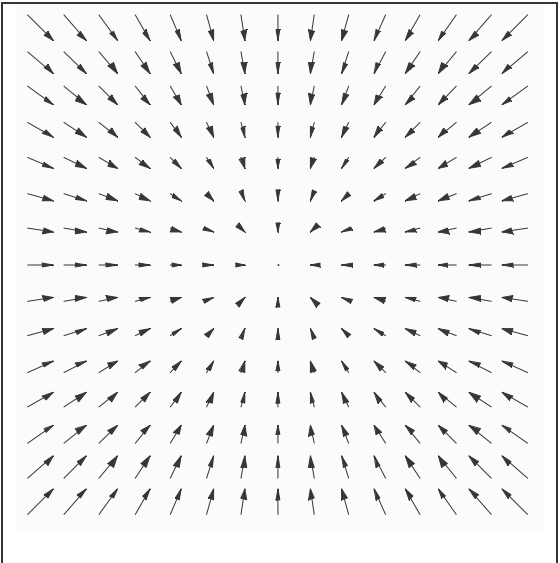 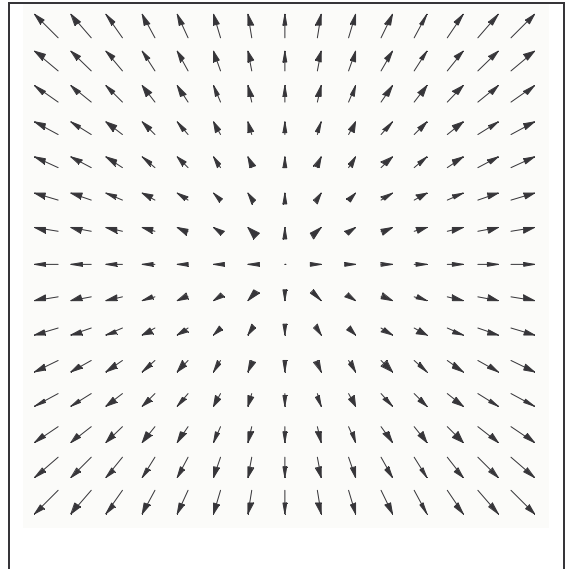

**Fig. 2:** The figure shows a large *positive* divergence as the arrows are pointing in.

**Fig. 3:** The figure shows a large *negative* divergence as the arrows are pointing out

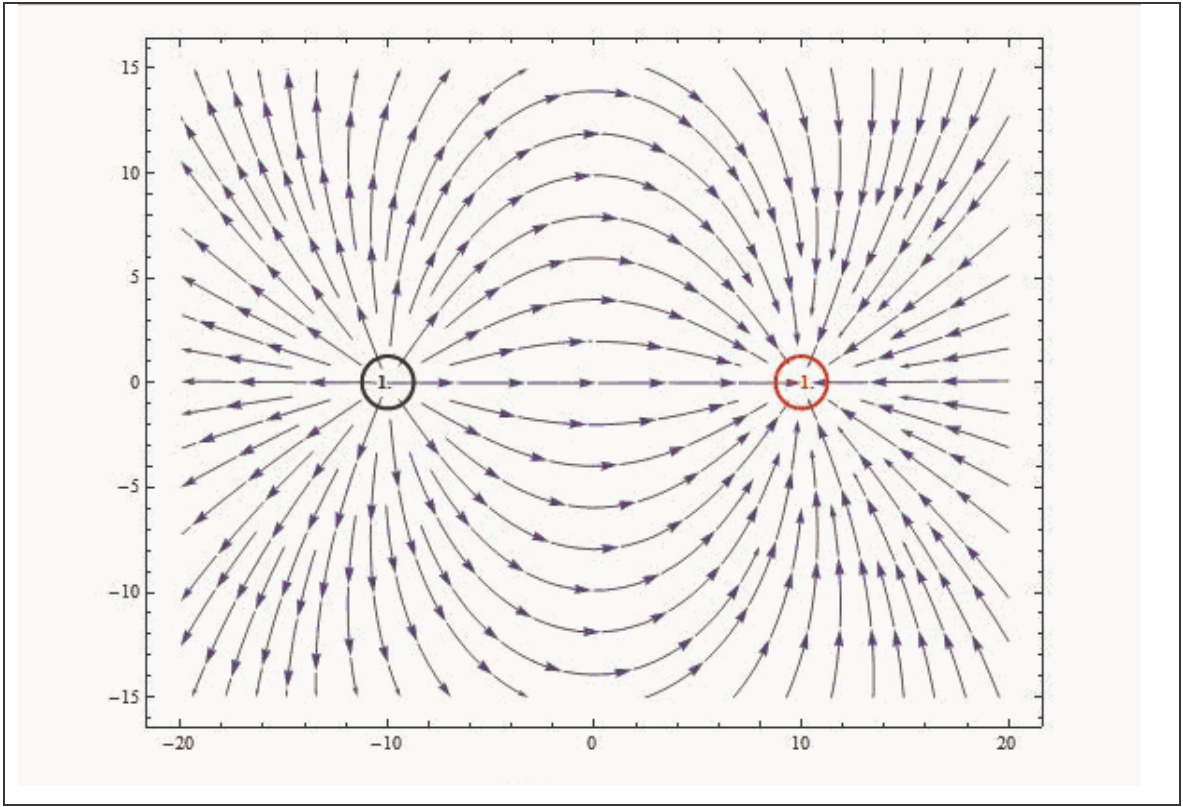

**Fig. 4:** A computer simulation of solenoidal nature of *probability flow* or *flow of the probability current*, on arbitrary scale (*courtesy*- Wolfram *Mathematica*).

The time independent character of the quantum states leads to

$$\left(\text{Div } \vec{S}\right) = 0. \tag{7}$$

This implies that the *Probability flow* (in analogy with fluid dynamics) or *flow of the Probability current* is solenoidal as depicted in the **Figure 4**. This is yet another manifestation of the underlying idea of Faraday loops [13].

An appropriate example in the sight is the case of theory of Loop Quantum Gravity (LQG), where the states and therefore the scalar product [27, 29] are time independent. In a modern perspective such Faraday loops are realised as the gauge invariant degrees of freedom for General Relativity and Yang-Mills theory in Loop Quantum Gravity (LQG) [14, 22, 27]. We give yet another interpretation to this idea as follows: The (Faraday) lines that make these loops are the *locus of probability flow*. Even if we factor away the diffeomorphism invariance, physical structures and dynamics, the abstract graph structure remains there [24- 26]. Thus, the hypothesis of graphs and loops even in the absence of physical objects or dynamics makes sense in terms of probability and probable paths. This is also motivation for us to explore these ideas in probability space.

**4. Statistical Distance in the Probability Space**

Having inspired by the idea of probability flow and loops, we explore the geometry of probability space. In Statistics, the probability distribution for the outcome of a physical process is given by $n+1$ real numbers $p_i$ such that

$$p_i \geq 0, \text{ and } \sum_{i=0}^{n} p_i = 1 . \tag{8}$$

The geometry of the $n$ - sphere is almost manifest here [9], and we find it such that

$$\xi_i \equiv \sqrt{p_i} \implies \sum_{i=0}^{n} \xi_i^2 = 1 . \tag{9}$$

We see here that the space of all probability distributions lies on a sphere embedded in a flat space [9], and therefore it carries the natural metric of the sphere, namely:

$$ds^2 = \sum_{i=0}^{n} d\xi_i d\xi_i = \frac{1}{4} \sum_{i=0}^{n} \frac{dp_i dp_i}{p_i} . \tag{10}$$

In Statistics this is known as the Bhattacharya metric or Fisher-Rao metric [9] (see references [10], [11], and [31] also). It enables us to define the geodesic distance between two arbitrary probability distributions; if we consider the case where there are only two possible outcomes then the geodesic distance between two probability distributions $(1-p_1, p_1)$ and $(1-p_2, p_2)$ is:

$$\cos d = \sqrt{(1-p_1)(1-p_2)} + \sqrt{p_1 p_2} \,. \tag{11}$$

Therefore distance $d$ is given by:

$$d = \cos^{-1}(\sqrt{(1-p_1)(1-p_2)} + \sqrt{p_1 p_2}) \,. \tag{12}$$

To explore further geometry of this expression we substitute here

$p_1 = x^2$, and $p_2 = y^2$, and find that

$$d = \cos^{-1}(\sqrt{(1-x^2)(1-y^2)} + xy) \,. \tag{13}$$

Using the identity of inverse circular functions we find that-

$$\cos^{-1}(\sqrt{(1-x^2)(1-y^2)} + xy) = \cos^{-1}(x) - \cos^{-1}(y) \,; \tag{14}$$

Therefore, the expression of geodesic in probability space is equivalent to

$$d = \cos^{-1}(\sqrt{p_1}) - \cos^{-1}(\sqrt{p_2}) \,. \tag{15}$$

The graphical plot of this geodesic in equation (15) appears as follows:

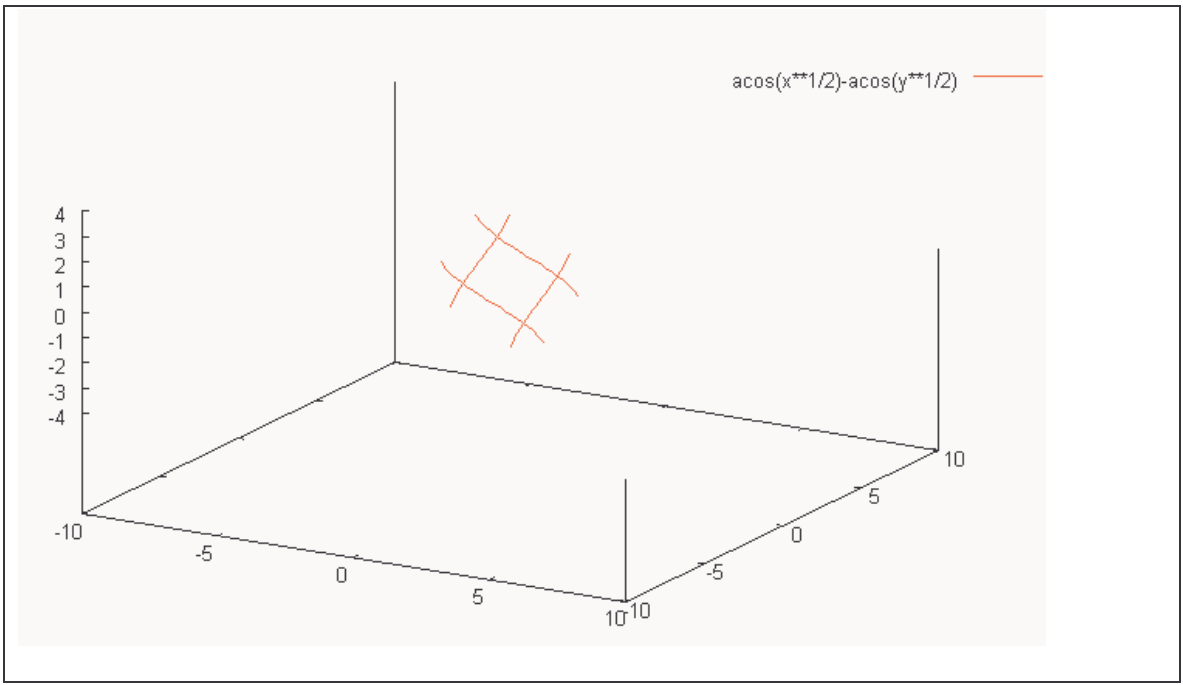

**Fig. 5**: Graph representing geodesic $d = \cos^{-1}(\sqrt{p_1}) - \cos^{-1}(\sqrt{p_2})$ in probability space.

This is precisely a loop form. The joints of the loop imply the possibilities of further loop attachments with this loop and so on. And ultimately that may result into a mesh or grid. Also, it is interesting to note that abstract graph structure is not directed one. It is only in a specific physical framework that these graphs appear to be directed.

**5. Eigen-Values of the Area Operator and Geometry of the Corresponding Surfaces**

Regulated operators corresponding to areas of 2-surfaces have been introduced [4, 24, 25, 28, 29] and shown to be self- adjoint on the underlying (kinematical) Hilbert space of states. It is shown that the spectra corresponding to these area operators are purely discrete. There is indication that underlying quantum geometry is far from the continuum picture. In fact, the fundamental excitations of quantum geometry are one-dimensional, and the three- dimensional continuum geometry emerges only on the coarse graining.

The physical area $A$ of a surface $\Sigma$ depends on the metric, namely the gravitational field. In a quantum theory of gravity, the gravitational field is a quantum field operator, and therefore we must describe the area of $\Sigma$ in terms of a quantum observable, described by an operator $\hat{A}$. This can be worked out by writing the standard formula for the area of a surface, and replacing the metric with the appropriate function of the loop variables. Promoting the loop variables to operators, we obtain the area operator $\hat{A}$.

The resulting area operator $\hat{A}$ acts on a spin network state $|S\rangle$ (assuming for simplicity that $S$ is a spin network without nodes on $\Sigma$) as follows:

$$\hat{A}(\Sigma)|S\rangle = \left( \frac{l_0^2}{2} \sum_{i \in \{S \cap \Sigma\}} \sqrt{p_i(p_i + 2)} \right) |S\rangle. \tag{16}$$

Where,

$$l_0^2 = \hbar G = \frac{16\pi\hbar G_{Newton}}{c^3} = 16\pi l_{Planck}^2 ; \tag{17}$$

$i$ labels the intersections between the spin network $S$ and the surface $\Sigma$, and $p_i$ is the color of the link of $S$ crossing the $i$-th intersection. This result shows that the spin network states are eigen states of the area operator. The corresponding spectrum is labeled by multiplets $p_i = (p_1,......,p_n)$ of positive half integers, with arbitrary $n$, and is given by

$$a_{p_i}(\Sigma) = \left(\frac{l_0^2}{2} \sum_i \sqrt{p_i(p_i+2)}\right). \tag{18}$$

Shifting from color to spin notation reveals the $SU(2)$ origin of the spectrum with the eigen-values:

$$a_S = \frac{l_{Pl}^2}{2}\sqrt{j(j+1)} . \tag{19}$$

If we choose $j = \frac{1}{2}$, we obtain the eigen-values $a_S^0$ of three types as follows:

$$a_S^0 = \frac{\sqrt{3}}{4} l_{Pl}^2 , \tag{20}$$

$$a_S^0 = \frac{2\sqrt{2}}{4} l_{Pl}^2 , \text{ and} \tag{21}$$

$$a_S^0 = \frac{2}{4} l_{Pl}^2 . \tag{22}$$

Now a few vital questions arise: How does this area appear to be? What surface it represents? If it is an oriented area, where and how it is oriented?

To answer these questions we explore the geometric possibilities of representing these eigen-values of area. On careful examination of the expression of area eigen- values we make certain observations. It represents neither a surface on a plane nor does it lie on a spherical surface.

Having ruled out these possibilities, we explore other possibilities. We argue that an underlying surface being represented by the area operator must be surface of revolution. The term $8\pi l_{Pl}^2$ signifies ordinary geometric area of a surface. The only distinct feature of these expressions of the area spectrum is the term- $\sqrt{j(j+1)}$. Here we get a clue from spectroscopy. The term $\sqrt{j(j+1)}$ is reminiscent of the well-known $S-S$ coupling in atomic spectroscopy [30]. Coincidently, our areal spectra emerges out of spin-network states where $S-S$ coupling is very much part of it. Thus, if we plot the surface of revolution that has genesis in $S-S$ coupling, the following pictures emerge:

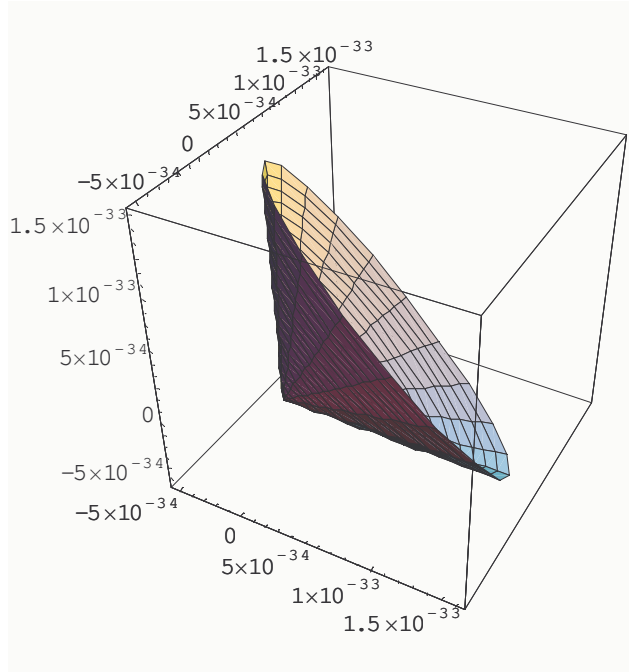

**Fig. 6**: Surface of revolution corresponding to area: $8\pi l_{Pl}^2$ ; Revolution Axis $\rightarrow \{1,1,1\}$.

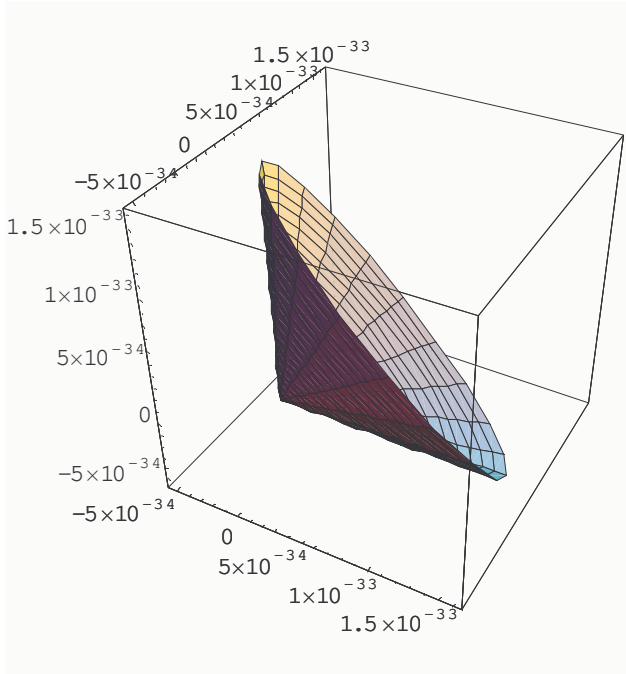

**Fig. 7:** Surface of revolution corresponding to area: $\dfrac{\sqrt{3}}{4} 8\pi l_{Pl}^2$. Revolution Axis $\to \{1,1,1\}$.

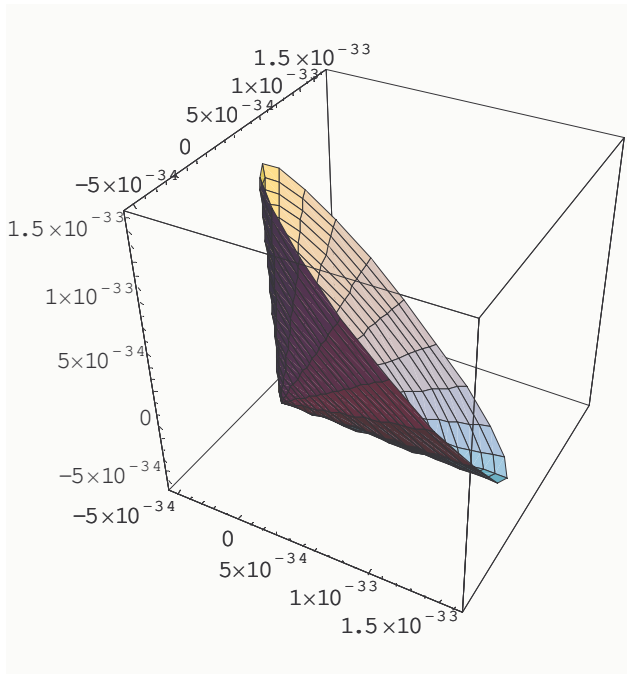

**Fig. 8:** Surface of revolution corresponding to area: $\dfrac{2}{4}(8\pi l_{Pl}^2)$. Revolution Axis $\to \{1,1,1\}$.

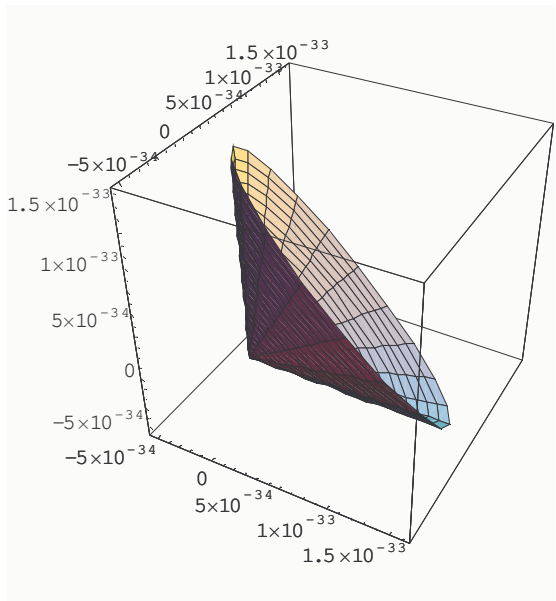

**Fig. 9:** Surface of revolution corresponding to area: $\dfrac{2\sqrt{2}}{4}(8\pi l_{Pl}^{2})$. Revolution Axis $\to \{1,1,1\}$.

In atomic spectroscopy [30], the surface of revolution is the cone traced out by $S$, whereas the atom is sitting at the tip of the cone. If we set the limits of the parameter for surface of revolution to $\to \{-l_{Pl}, l_{Pl}\}$, then the same amount of area could be observed as two surfaces of revolution connected as shown in the following figures:

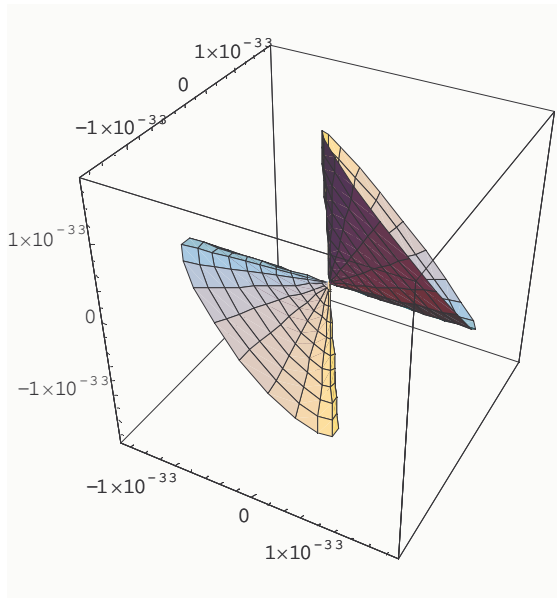

**Fig. 10:** Surface of revolution corresponding to area: $8\pi l_{Pl}^{2}$ ;

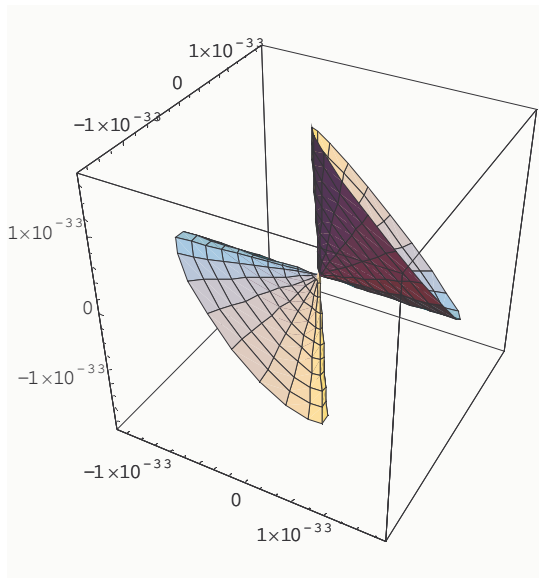

**Fig. 11:** Surface of revolution corresponding to area: $\dfrac{\sqrt{3}}{4} 8\pi d_{Pl}^2$ ;

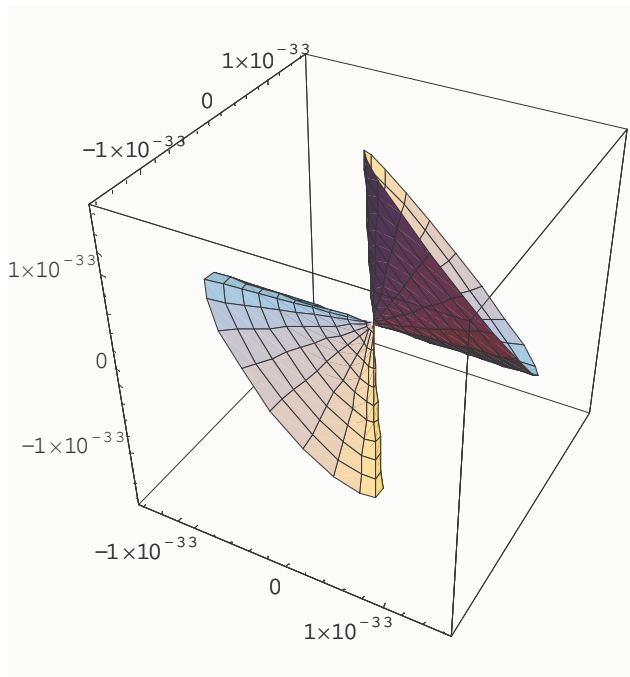

**Fig. 12:** Surface of revolution corresponding to area: $\dfrac{2}{4}(8\pi d_{Pl}^2)$ ;

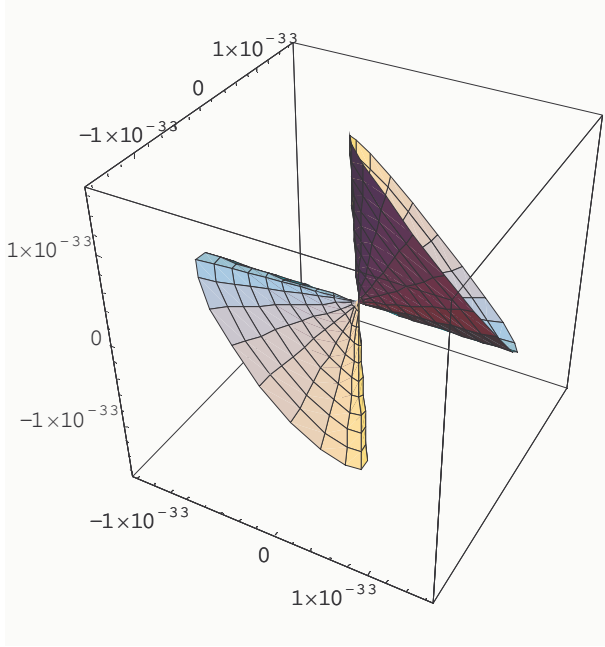

**Fig. 13:** Surface of revolution corresponding to area: $\dfrac{2\sqrt{2}}{4}(8\pi l_{Pl}^2)$;

Obviously, these area graphics represent a simple picture. And $S-S$ coupling among multiple spin states is yet to be explored. This is the limitation of the exercises in this discussion.

**6. Summary and Discussion**

However, the graphics created for picturisation of the eigen-values of area operators represent a simpler aspect of the problem. The $S-S$ coupling among multiple spin states is a much more complicated scenario. It is considered to be complicated in atomic spectroscopy too [30]. And exploration of such a coupling and its geometrical implications for spin-network with large number of spin states is mooted as an open problem. Thus further exploration to build up a complete picture is desirable.

The present discussion is aimed at emphasizing the importance of idea of probability flow and loops therein. This in turn further motivates us for more explorations in the probability space. We suggest the creation of graphics for binary space, and particularly probability space as further explorations. This may result into a powerful tool for exploring theory of *Quantum Computation* and also *Quantum Information Theory*.

§ The graphic in Fig. 1 is created by G*nuPlot*, the ditto plot is obtained by *MatLab*.

⊗ The graphic in Fig. 5 has been created using G*nuPlot*.

\* Computer simulations in Figures 6 -12 have been generated using *Mathematica*.

## Acknowledgements

The tips given by Wolfram *Mathematica* for Graphic in Figure 4 are gratefully acknowledged. Also, the author wishes to thank Prof. Sardar Singh for explaining niceties of Spectroscopy.